\documentclass[11pt]{iopart}

\usepackage{amsgen}
\usepackage{amssymb}
\usepackage{graphicx}
\usepackage{setstack}
\usepackage{color}
\usepackage{hyperref}

\begin{document}

\title{Gravitational Slip in the Parameterized Post-Newtonian Cosmology}

\author{Theodore Anton$^{1a}$, Timothy Clifton$^{2a}$\footnote{corresponding author} and Daniel B. Thomas$^{3b}$}
\address{$^a$Department of Physics \& Astronomy, Queen Mary University of London, UK.\\
$^b$Jodrell Bank Centre for Astrophysics, The University of Manchester, UK.}
\ead{$^1$theoanton123@gmail.com, $^2$t.clifton@qmul.ac.uk, $^3$dan.b.thomas1@gmail.com}

\begin{abstract}

A key signature of general relativity is that the two scalar potentials $\Phi$ and $\Psi$, when expressed in the longitudinal gauge, are equal in the absence of fluids with anisotropic stress. This is often expressed by stating that their ratio, the ``gravitational slip'', is equal to unity. However, the equality of $\Phi$ and $\Psi$ is typically broken in alternative theories of gravity. Observational constraints on the slip parameter are therefore of direct interest for testing Einstein's theory. In this paper we derive theory-independent expressions for the slip parameter on both large and small scales in Friedmann cosmologies, expressing it as a function of the post-Newtonian parameters. This is the final ingredient required for a complete parameterization of dust and dark energy-dominated cosmologies within the framework of Parameterized Post-Newtonian Cosmology (PPNC), which allows for the fully self-consistent modelling of cosmological observables without assuming any specific theory of gravity.

\end{abstract}

\vspace{-0.5cm}
\section{Introduction}\label{sec:intro}

Testing relativistic theories of gravity using cosmological data has a long history, and is a key science goal of many observational missions including Euclid \cite{euclid}, the SKA \cite{ska} and the Rubin Observatory \cite{lsst}. Given the enormous resources being invested into these facilities, it is of the utmost importance to ensure that theoretical frameworks are in place that can be applied to maximally exploit the data that will result from them \cite{pedro}. Recent proposals for such frameworks are, however, often restricted in the classes of modified gravity theories that they encompass, and can also be restricted by the introduction of additional assumptions, such as the imposition of a $\Lambda$CDM background cosmology. Although such approaches have their merits, we expect that it will also be helpful for the cosmology community to have at its disposal an approach that makes fewer requirements on the underlying theory, and that consistently treats both background and perturbations. In this paper we contribute to this goal by providing the final part of a theory-independent framework that we have dubbed ``Parameterized Post-Newtonian Cosmology'' (PPNC) \cite{ppnc1}-\cite{ppnc5}. The guiding philosophy of the PPNC approach is to remain theory agnostic, while maintaining theoretical self-consistency and keeping the number of free parameters to a minimum. 

The PPNC formalism builds on the great successes of the Parameterized Post-Newtonian (PPN) framework \cite{tegp}, which is designed to be applied for similar purposes in isolated astrophysical systems on small scales. It parameterizes the {  scalar \cite{ppnc1,ppnc2,ppnc3} and vector \cite{ppnc3}} degrees of freedom that appear in the metric\footnote{{ Gravitational tensor modes couple to matter at quite high orders in post-Newtonian expansions, and so cannot be included as naturally as scalars and vectors.}}, without reference to any additional fields or prior geometric quantities that might be postulated in specific theories of gravity. This is achieved by constructing a global cosmology from gluing together a large number of sub-horizon-sized regions of space-time, each of which is individually well-described by the PPN ``test metric'' \cite{ppnc1}. The result is a set of Friedmann-like equations, together with linear perturbation equations, that are all simultaneously characterized by a single set of gravitational coupling parameters: the PPNC parameters. This approach to cosmology is not only consistent with the approach used in the PPN framework, but is explicitly isometric to the PPN test metric within any small region of space and time.

To date, we have been able to construct all aspects of a Friedmann cosmology filled with dust and dark energy, including the gravitational fields of non-linear structures, with the exception of a large-scale equation for the {\it gravitational slip} \cite{firstslip}. This missing part of the puzzle corresponds to a relationship between the scalar gravitational potentials $\Phi$ and $\Psi$ in the longitudinal (or Poisson) gauge, and can be taken to be of the form \cite{PPF}
\begin{equation} \label{slipgeneral}
\Phi = \eta\,\Psi \,,
\end{equation}
where $\eta$ is some function of time and spatial scale. For the case of general relativity it is well-known that Einstein's equations with negligible anisotropic stress imply that we have simply $\Phi=\Psi$. For alternative theories of gravity this relationship can be, and in general is, substantially different.

It is the goal of this paper to derive a theory-independent expression for the slip in the form of Equation (\ref{slipgeneral}), to be used on all cosmological scales and (to the extent that it is possible) in terms of the existing PPNC parameters only. We will find that it is necessary to make a number of assumptions in order to make this problem tractable. We do this by implicitly defining an effective stress-energy tensor $T_{\mu \nu}^{\rm eff}$ via the relationship $G_{\mu \nu} = 8 \pi \, T_{\mu \nu}^{\rm eff}$.
We can then define an effective anisotropic stress tensor $\Pi^{\rm eff}_{\mu \nu} = T^{\rm eff}_{\: \langle \mu\nu\rangle} = T^{\rm eff}_{\rho \sigma} \left(h^{\rho}_{\phantom{\rho}(\mu}h^{\sigma}_{\phantom{\sigma}\nu)} - \frac{1}{3}h_{\mu\nu}\,h^{\rho\sigma}\right)$, where $h_{\mu \nu} = g_{\mu \nu} +u_{\mu} u_{\nu}$ is a spatial projection tensor onto the spaces orthogonal to the time-like unit vector $u^{\mu}$, and angled brackets denote the projected, symmetric and trace-free part of a tensor. We will then (at different points) suppose:
\begin{itemize}
\item[] \hypertarget{assumption_1}{(1) Dark energy does not source $\Pi^{\rm eff}_{\mu \nu}$.}
\item[] \hypertarget{assumption_2}{(2) Dark energy is not sourced by $\Pi^{\rm eff}_{\mu \nu}$.}
\end{itemize}
These assumptions both aim to disassociate dark energy from effective anisotropic stress, but go in different directions. Both are satisfied if anisotropic stress and dark energy are completely unrelated, as is usually assumed in cosmology. We will elucidate upon them further as we progress through the paper.

This will progress as follows: in section \ref{sec:ppnc} we will recap the essential features of the PPNC approach to cosmological modelling. In section \ref{sec:bianchi} we will then consider the effective large-scale anisotropic stress produced by homogeneous anisotropic cosmologies of Bianchi type I, in order to relate gravitational tidal forces on large scales to those on small scales. We will use this result in section \ref{sec:direct} in order to derive an expression for the gravitational slip parameter. This is followed by the consideration of canonical scalar-tensor theories as an example that fits into our formalism in section \ref{sec:st}, followed by the results of numerically implementing our equations in the CLASS Boltzmann code, and applying them to the Planck $TT$, $TE$, $EE$ and lensing data, in section \ref{sec:num}. Finally, we summarize our findings in section \ref{sec:disc}, and present an alternative expression for the slip in Appendix A and a discussion of scale dependence in Appendix B. The principal results of each section are presented in boxes.

\section{Parameterized Post-Newtonian Cosmology} \label{sec:ppnc}

The PPNC formalism can be written in terms of a single {  spatially-flat} Friedmann-Lema\^{i}tre-Robertson-Walker cosmology as\footnote{{  Spatial curvature can also be included if desired \cite{ppnc1}, though we have neglected it here as it does not affect our current considerations.}}
\begin{equation}
{\rm d}s^2 = -\left( 1-2 \Phi \right) {\rm d} t^2 + a(t)^2 \, \left( 1+ 2 \Psi \right) \, \delta_{ij} \, {\rm d} x^i {\rm d}x^j \, ,
\end{equation}
where $a(t)$ is the scale factor, and where $\Phi$ and $\Psi$ are linear-order scalar perturbations to the otherwise perfectly homogeneous and isotropic background. We have chosen here to write the geometry in longitudinal gauge, and to neglect vector and tensor perturbations, as well as higher-order scalar perturbations and spatial curvature in the background.

\vspace{15pt}
\noindent
{\it Background Cosmology}
\vspace{5pt}

\noindent
At zeroth-order in perturbations we find the generalized Friedmann equations are
\begin{eqnarray} \label{f1}
\frac{\dot{a}^2}{a^2} &=& \frac{8 \pi}{3} \, \gamma \, \bar{\rho} - \frac{2 \gamma_c}{3} \\[5pt] \label{f2}
\frac{\ddot{a}}{a} &=&  -\frac{4 \pi}{3} \, \alpha \, \bar{\rho} + \frac{\alpha_c}{3} \, ,
\end{eqnarray}
which come together with the energy conservation equation 
\begin{equation}
\dot{\bar{\rho}} + 3 \frac{\dot{a}}{a} \bar{\rho}=0 \, ,
\end{equation}
where we have expanded the energy density of matter as $\rho = \bar{\rho} + \delta \rho$, such that $\bar{\rho}$ is the background value, and where we have assumed that radiation is negligible at late times. The parameters $\alpha=\alpha(t)$ and $\gamma =\gamma(t)$ are generalisations of the PPN parameters given by the same symbols, and which reduce to those parameters for an observer at the cosmological time $t$. The new parameters $\alpha_c=\alpha_c(t)$ and $\gamma_c=\gamma_c(t)$ are negligible in small-scale astrophysical systems such as the Solar System and binary pulsars, but are required to account for an homogeneous field of dark energy in the cosmological context. They are required to obey the integrability condition
\begin{equation} \label{intcon}
\alpha_c + 2 \gamma_c + \gamma_c' = 4 \pi \bar{\rho} \left( \alpha - \gamma + \gamma' \right) \, ,
\end{equation}
where a prime denotes differentiation with respect to the number of e-foldings of expansion, such that $X' = {\rm d} X/ {\rm d} \ln a$.

\vspace{15pt}
\noindent
{\it Scalar Perturbations}
\vspace{5pt}

\noindent
The leading-order PPNC scalar perturbation equations are 
\begin{eqnarray} \label{Hcon}
&&{H} \dot{\Psi} +  {H}^2 \Phi -\frac{1}{3 a^2} \nabla^2 \Psi = \frac{4 \pi}{3} \, \mu \, \delta \rho \, \\[5pt]
&&2 \left(H^2 +\dot{H}\right) \Phi +{H} \dot{\Phi}+\ddot{\Psi}+2 {H} \dot{\Psi} + \frac{1}{3 a^2} \nabla^2 \Phi = -\frac{4 \pi}{3} \, \nu \, \delta \rho \\[5pt]
\label{q} &&\dot{\Psi}_{,i} + H \Phi_{,i} = 4 \pi \, \mu \, 
\rho \, v_i \, a^2 
+ \mathcal{G} \, H \, \Psi_{,i} \, ,
\end{eqnarray}
together with the energy conservation and Euler equations
\begin{eqnarray}
&&(\delta \rho{)}^{\cdot} + 3 H \, \delta \rho + 3 \bar{\rho} \, \dot{\Psi} = 0 \\[5pt]
&&\dot{v}_i + 2H\,v_i - \frac{1}{a^2}\Phi_{,i} = 0\,,
\end{eqnarray}
where $H=\dot{a}/a$ is the Hubble rate, $\nabla^2$ is the three-dimensional spatial Laplacian, and $v_i$ is the coordinate 3-velocity of matter. The parameters $\mu$, $\nu$ and $\mathcal{G}$ have limits that can be written in terms of the PPN parameters as \cite{ppnc2, ppnc3}
\vspace{5pt}
\begin{center}
\begin{tabular}{ l l }
 $\displaystyle \lim_{L\rightarrow 0} \mu= \gamma$ & $\displaystyle\lim_{L\rightarrow \infty} \mu= \gamma - \frac{1}{3} \gamma' + \frac{1}{12 \pi \bar{\rho}} \gamma_c'$  \\ 
 $\displaystyle \lim_{L\rightarrow 0} \nu= \alpha$ & $\displaystyle \lim_{L\rightarrow \infty} \nu = \alpha - \frac{1}{3} \alpha' + \frac{1}{12 \pi \bar{\rho}} \alpha_c'$  \\  
$\displaystyle \lim_{L\rightarrow 0} \mathcal{G} = \frac{\alpha-\gamma}{\gamma} +\frac{\gamma'}{\gamma} \hspace{2cm}$ & $\displaystyle \lim_{L\rightarrow \infty} \mathcal{G} = 0 \, ,$     
\end{tabular}
\end{center}
\vspace{0pt}

\noindent
where primes again denote derivatives with respect to e-foldings of expansion {  and where $L$ is the length scale of the system under consideration}. For a more detailed discussion of the underlying principles of the formalism, we refer the reader to references \cite{ppnc1}-\cite{ppnc5}.

\vspace{15pt}
\noindent
{\it The Missing Slip}
\vspace{5pt}

\noindent
Missing from the equations above is a relationship of the form given in Equation (\ref{slipgeneral}), specifying the relationship between $\Phi$ and $\Psi$. On small scales, where post-Newtonian expansion can be applied, {  we have $\nabla^2 \Phi = - 4 \pi \, \alpha \, a^2 \delta \rho$ and $\nabla^2 \Psi = - 4 \pi \, \gamma \, a^2 \delta \rho$ \cite{ppnc1}, which (assuming similar boundary conditions) gives the slip} in terms of PPN parameters as
\begin{equation} \label{ssslip}
\boxed{
\Phi = \frac{\alpha}{\gamma} \, \Psi } \, .
\end{equation}
On large scales, however, the corresponding expression is harder to find. We note that such an equation would normally be interpreted as a constraint, though in the full set of Einstein equations it would actually result from the shear evolution equation (or the projected symmetric tracefree (PSTF) part of the Gauss embedding equation). It is the choice of longitudinal gauge that is responsible for setting the shear to zero, and which in Einstein's equations then results in the equation $\Phi=\Psi$, in the absence of anisotropic stress. The slip therefore has a different character to the other constraint equations, which are constraints by virtue of their role in the full, non-perturbative Einstein equations, rather than being enforced by a choice of gauge in cosmological perturbation theory. In fact, it is straightforward to verify that the Hamiltonian and momentum constraint equations (\ref{Hcon}) and (\ref{q})  can be consistently evolved by the evolution equations {\it without} any additional information on a relationship between $\Phi$ and $\Psi$ (i.e. without a ``slip'' equation). This makes the slip a dynamical constraint, which can be freely specified at every moment of time without interfering with the consistency of the equations above. Correspondingly, this also suggests that the slip may come from elsewhere within the context of our theory-independent approach: we may expect it to arise from a dynamical evolution equation, rather than a constraint on any individual constant-$t$ hypersurface. We will investigate how this works in the following sections.

\section{Large-Scale Slip Equations as a PSTF Perturbation Equation}\label{sec:bianchi}

All aspects of the scalar and vector parts of large-scale perturbations to Friedmann have already been calculated \cite{ppnc1}-\cite{ppnc5}. We are therefore led to the projected symmetric tracefree part of large-scale perturbations as a possible source for the slip equation. This possibility is supported by the slip equation arising from the corresponding sector in Einstein's equations, as well as this being the next simplest type of perturbation that one could consider.

\subsection{Theory-Independent Bianchi I Cosmologies}

In order to obtain a non-trivial PSTF perturbation of Friedmann on large scales, we can use a separate universe approach in which the comparison is made between an anisotropic Bianchi cosmology and an isotropic Friedmann cosmology. For this purpose, the simplest cosmologies are of Bianchi class I, with line-element
\begin{equation} \label{ds2}
{\rm d}s^2 = -{\rm d}t^2 +A^2(t) {\rm d}x^2 + B^2(t) {\rm d}y^2 + C^2(t) {\rm d}z^2 \, .
\end{equation}
This space-time has different expansion rates in the $x$, $y$ and $z$ directions. It can therefore accommodate non-trivial rank-2 PSTF equations, such as the shear evolution equation.

To determine the emergent field equations for this type of cosmological model requires us to relate the scale factors (and their derivatives) to the energy density of matter and the PPNC parameters. For this, we need to transform a small patch ($L \lesssim 100 \, {\rm Mpc}$) of the geometry described by the line-element in Equation (\ref{ds2}) into perturbed Minkowski space:
\begin{equation} \label{ds2b}
\hspace{-2cm}
ds^2 = - (1-2 {\phi}) \, d\hat{t}^2 + (1+2{\psi}_1) \, d\hat{x}^2 + (1+2{\psi}_2) \, d\hat{y}^2 + (1+2{\psi}_3) \, d\hat{z}^2 + O(\eta^3) \, ,
\end{equation}
where $\eta$ is the post-Newtonian expansion parameter, defined such that the time-derivative of a quantity is of order $\eta$ times smaller than its spatial derivative. This transformation can be achieved by implicitly defining the coordinates $\{\hat{t},\hat{x},\hat{y},\hat{z}\}$ on perturbed Minkowski space-time through the following equations:
\begin{eqnarray}
t = \hat{t} - \frac{1}{2} D
\qquad {\rm and} \qquad
x^i = \frac{\hat{x}^i}{a_i} \left( 1 + \frac{1}{4} \frac{\dot{a}_i}{a_i} D \right) 
\end{eqnarray}
where $a_i \in \{ A,B,C \}$ and $D \equiv  \frac{\dot{A}}{A} \, \hat{x}^2 + \frac{\dot{B}}{B} \, \hat{y}^2 + \frac{\dot{C}}{C} \, \hat{z}^2 $, and where no sums over $i$ are implied in the second equation. This gives the post-Newtonian gravitational potentials in Equation (\ref{ds2b}) as
\begin{eqnarray} \label{phi}
{\phi} &=& \frac{1}{2} \left( \frac{\ddot{A}}{A} \, \hat{x}^2 + \frac{\ddot{B}}{B} \, \hat{y}^2 + \frac{\ddot{C}}{C} \, \hat{z}^2 \right)
\qquad {\rm and} \qquad
{\psi}_i = -\frac{1}{4} \frac{\dot{a}_i}{a_i} D \, ,
\end{eqnarray}
where {  $\psi_i \in \{\psi_1,\psi_2,\psi_3\}$}, and where again no sums over $i$ are implied. We now wish to write the line-element in Equation (\ref{ds2b}) in standard post-Newtonian gauge, so that it appears as
\begin{equation} \label{ds2c}
ds^2 = - (1-2 \phi) \, d\hat{t}^2 + (1+2\psi) \left( d\hat{x}^2 +d\hat{y}^2 +d\hat{z}^2 \right) \, ,
\end{equation}
where the equality here is  up to transverse tracefree perturbations (which cannot be removed by gauge transformations). This can be achieved by performing a gauge transformation on the spatial coordinates such that $\hat{x}^i \rightarrow  \hat{x}^i + \xi^i$, where $\xi^i$ satisfies 
\begin{eqnarray}
\hat{\nabla}^2 \xi^{\hat{x}} &=& \frac{1}{2} \left( \frac{\dot{A}^2}{A^2}- \frac{\dot{B}}{B} \frac{\dot{C}}{C} \right) \hat{x} \, ,
\end{eqnarray} 
where $\hat{\nabla}$ is the Laplacian operator in the hatted coordinates, and where $\hat{\nabla}^2 \xi^{\hat{y}}$ and $\hat{\nabla}^2 \xi^{\hat{z}}$ are given by cyclic permutations. In this case the spatial perturbation reads
\begin{eqnarray} \label{psi}
\hspace{-2cm}
{\psi} = -\frac{1}{4} \left[\frac{\dot{A}}{A} \frac{\dot{B}}{B} (\hat{x}^2+\hat{y}^2-\hat{z}^2)
+\frac{\dot{B}}{B} \frac{\dot{C}}{C} (\hat{y}^2+\hat{z}^2-\hat{x}^2)
+\frac{\dot{C}}{C} \frac{\dot{A}}{A} (\hat{z}^2+\hat{x}^2-\hat{y}^2) \right] \,.
\end{eqnarray}
The line-element in Equation (\ref{ds2c}) is now in standard post-Newtonian gauge, with the metric perturbations $\phi$ and $\psi$ given by the expressions in equations (\ref{phi}) and (\ref{psi}).

Let us now recall the equations that relate the gravitational potentials to the mass density and PPNC parameters \cite{ppnc1}:
\begin{eqnarray}
\hat{\nabla}^2 \phi &=& -4 \pi \, \alpha \, \rho + \alpha_c \\
\hat{\nabla}^2 \psi &=& -4 \pi \, \gamma \, \rho + \gamma_c \, .
\end{eqnarray}
Using these together with equations (\ref{phi}) and (\ref{psi}) gives, after integration,
\begin{eqnarray}
&& \frac{\ddot{A}}{A} + \frac{\ddot{B}}{B} +\frac{\ddot{C}}{C} = -4 \pi \, \alpha \, {\rho} + \alpha_c \\
&& \frac{\dot{A}}{A} \frac{\dot{B}}{B} + \frac{\dot{B}}{B} \frac{\dot{C}}{C} +\frac{\dot{C}}{C} \frac{\dot{A}}{A} = 8 \pi \, \gamma \, {\rho} - 2 \gamma_c \, ,
\end{eqnarray}
where ${\rho}$ is homogeneous. These equations reduce to the emergent Friedmann equations (\ref{f1}) and (\ref{f2}) in the case where $A$, $B$ and $C$ are all equal (such that the space-time is isotropic). The energy conservation equation in this case is exactly the same as it always is in Bianchi I cosmologies:
\begin{equation}
\dot{{\rho}} + \left( \frac{\dot{A}}{A} + \frac{\dot{B}}{B} + \frac{\dot{C}}{C} \right) {\rho} = 0 \, .
\end{equation}
As well as the equations above, in the case of anisotropic cosmologies we also have the additional three equations below:
\begin{eqnarray}
\frac{1}{3} \left[ 2 \frac{\ddot{A}}{A} - \frac{\ddot{B}}{B} - \frac{\ddot{C}}{C} + \frac{\dot{A}}{A} \left( \frac{\dot{B}}{B} + \frac{\dot{C}}{C} \right) - 2 \frac{\dot{B}}{B} \frac{\dot{C}}{C} \right] 
=\left({\phi} - {\psi} \right)_{, \langle \hat{x}\hat{x} \rangle} 
\, ,
\end{eqnarray}
{  where the angle brackets around the derivatives indicate the symmetric and trace-free spatial derivative, defined such that e.g. $\phi_{,\langle i j \rangle} = \phi_{,(ij)} -\frac{1}{3} \delta_{ij} \phi_{,kk}$, and} with cyclic permutations for the $\hat{y}\hat{y}$ and $\hat{z}\hat{z}$-components. These three equations vanish in the case $A=B=C$. It remains to determine how to calculate the right-hand side in terms of PPNC and matter parameters.

\subsection{Bianchi I as a Perturbation of Robertson-Walker Geometry}

Having derived expressions for the Bianchi I equations in terms of small-scale post-Newtonian potentials, we now need to describe that same Bianchi I space-time as a perturbation of Friedmann in order to derive an equation for the large-scale slip. We can start with a spatially-flat Robertson-Walker geometry with scalar perturbations in longitudinal gauge \cite{bert}:
\begin{equation} \label{rw}
ds^2 = -(1-2 \Phi) dT^2 + a^2(T) (1+2 \Psi) \left(dX^2 +dY^2+dZ^2 \right) \,.
\end{equation}
Let us now change coordinates so that
\begin{equation}
T = t + \tau \qquad {\rm and} \qquad X^i = x^i + \zeta^i  \, ,
\end{equation}
where $\tau$ and $\zeta^i$ are small in the sense of cosmological perturbation theory, i.e. they are of the order of the perturbative expansion parameter $\epsilon$. Under this transformation the squared scale factor transforms as
\begin{equation}
a^2 (T) = a^2 (t) (1 + 2 H \tau) + O(\epsilon^2) \, ,
\end{equation}
and the squares of coordinate infinitessimals transform as
\begin{eqnarray} \label{dspert}
&&dT^2 = (1+2 \dot{\tau} )dt^2 +2\tau_{,i} \, dt \, dx^i  + O(\epsilon^2)\\
&&(dX^i)^2 = (dx^i)^2 +2 \dot{\zeta}_i \, dx^i dt + 2 \zeta_{i,j}\, dx^i dx^j+ O(\epsilon^2) \, .
\end{eqnarray}
Putting these transformations back into the perturbed Robertson-Walker geometry in Equation (\ref{rw}) gives
\begin{eqnarray}
ds^2 &=& -\left(1+2 \dot{\tau}-2 \Phi \right) dt^2 - 2 (\tau_{,i}- a^2 \dot{\zeta}_i) dx^i dt \nonumber \\
&&+ a^2 \left[ (1 +2 H \tau +2 \Psi)  \delta_{ij} +2 \zeta_{i,j} \right]dx^i dx^j +O(\epsilon^2) \, . \label{dspert2}
\end{eqnarray}
For this to be equivalent to the line-element of Bianchi I geometry, as shown in Equation (\ref{ds2}), requires all off-diagonal components of the spatial part of the metric to vanish, which implies
\begin{equation}
\zeta_{x,y} = \zeta_{x,z} = \zeta_{y,x}=\zeta_{y,z} = \zeta_{z,x} = \zeta_{z,y}=0 \,.
\end{equation}
In other words, any of the $\zeta^i$ can only be a function of the corresponding $x^i$ and $t$. To go further requires treating the scale factors $A$, $B$ and $C$ as perturbations of the isotropic scale factor $a$. We do this such that
\begin{equation} \label{abc}
A= a ( 1 +\delta_1) \, , \quad B=a (1+\delta_2) \quad {\rm and} \quad C= a(1+\delta_3) \, ,
\end{equation}
where $\delta_i \sim \epsilon$ and such that $\delta_1+\delta_2+\delta_3=0$. This means that the trace of the spatial part of the metric in Equation (\ref{dspert2}) should vanish, and the deviations from isotropy must be found within $\zeta_{i,j}$. This implies that $\Psi= - H \tau$, and that
\begin{equation}
\zeta_{x,x}=  \delta_1 \, , \quad \zeta_{y,y} = \delta_2 \quad {\rm and} \quad \zeta_{z,z} =  \delta_3 \, .
\end{equation}
We also require $\tau_{,i} = a^2 \dot{\zeta}_i$ for the time-space components of the metric to vanish, which gives us
\begin{equation}
\tau = \frac{a^2}{2} \left( \dot{\delta}_1 \, x^2 + \dot{\delta}_2 \, y^2 + \dot{\delta}_3 \, z^2 \right) \, .
\end{equation}
Finally, we can see that we must require $\Phi = \dot{\tau}$ for the time-time component to take the required form. Putting this all together gives us the following perturbations, for the Robertson-Walker geometry to be an approximation to the Bianchi I geometry to the required order:
\begin{eqnarray}
\Phi &=& \frac{a^2}{2} \left[ \left( 2 H \dot{\delta}_1 + \ddot{\delta}_1 \right) X^2 +  \left( 2 H \dot{\delta}_2 + \ddot{\delta}_2 \right) Y^2 +  \left( 2 H \dot{\delta}_3 + \ddot{\delta}_3 \right) Z^2 \right] \\
\Psi &=& -\frac{a^2}{2} H \left[ \dot{\delta}_1 \, X^2 + \dot{\delta}_2 \, Y^2 + \dot{\delta}_3 \, Z^2 \right] \, ,
\end{eqnarray}
where the $\delta_i$ are understood to have the meaning of the anisotropic part of each of the Bianchi scale factors, as follows from their definition in Equation (\ref{abc}). This gives us
\begin{equation}
\left( \Phi- \Psi \right)_{,\langle ii \rangle} = a^2 \left( 3H\, \dot{\delta}_i + \ddot{\delta}_i \right) \, ,
\end{equation}
where no sum is implied by the repeated index on the left-hand side. To the required accuracy, this can also be written {  in the following way (as can be verified by substitution)}
\begin{equation}
\hspace{-1cm}
\left( \Phi- \Psi \right)_{,\langle XX \rangle} = \frac{a^2}{3} \left[ 2 \frac{\ddot{A}}{A} - \frac{\ddot{B}}{B} - \frac{\ddot{C}}{C} + \frac{\dot{A}}{A} \left( \frac{\dot{B}}{B} + \frac{\dot{C}}{C} \right) - 2 \frac{\dot{B}}{B} \frac{\dot{C}}{C} \right] + O(\epsilon^2) \, ,
\end{equation}
{  with the $O(\epsilon^2)$ denoting all terms that are higher than linear order in $\delta_i$, and} with cyclic permutations for $\left( \Phi- \Psi \right)_{,\langle YY \rangle}$ and $\left( \Phi- \Psi \right)_{,\langle ZZ \rangle}$. {  This means}
\begin{equation} \label{ls1}
\hspace{-1cm}\boxed{
\big( \Phi- \Psi \big)_{,\langle ii \rangle} = \big( \hat{\Phi} - \hat{\Psi} \big)_{,\langle ii \rangle} + O(\epsilon^2)} \, , 
\end{equation}
where no sum over $i$ is implied, and where $\hat{\Phi}$ and $\hat{\Psi}$ are the small-scale potentials from Equation (\ref{ds2c}) but written in expanding coordinates. They are given explicitly by \cite{ppnc1}
\begin{eqnarray} \label{c1eq}
\hspace{-1cm}\hat{\Phi} &\equiv& \phi - \frac{1}{2} \frac{\ddot{a}}{a} \left(x^2+y^2+z^2 \right) \qquad {\rm and} \qquad
\hat{\Psi} \equiv \psi + \frac{1}{4} \frac{\dot{a}^2}{a^2} \left( x^2 + y^2 + z^2 \right) \, .
\end{eqnarray}
Equation (\ref{ls1}) is a relationship between large- and small-scale gravitational potentials, $\{\Phi, \Psi\}$ and $\{ \hat{\Phi}, \hat{\Psi}\}$. It shows they are closely related, despite having very different constraint and evolution equations. We note that there exists a residual freedom, on large scales, to rescale spatial coordinates by $X^i \rightarrow X^i (1-c_1)$ for any constant $c_1$, resulting in $\Psi \rightarrow \Psi - c_1$. This will be used later on{\color{red}, and is discussed further in Appendix A}. The result in Equation (\ref{ls1}) is significant because it tells us that the effective anisotropic stress that emerges from small scales will precisely control the effective anisotropic stress (and therefore the gravitational slip) on super-horizon scales. With this result in hand, we can now calculate that emergent stress, in the knowledge that it will directly result in the large-scale slip.

\section{Constructing the Large-Scale Slip} \label{sec:direct}

%Although Equation (\ref{effcon}) is already applicable to a wide range of gravity theories, it would be particularly helpful to write the slip purely in terms of the PPNC parameters, independently of the Hubble rate. In order to do this, we require a direct, theory-agnostic, construction of the large-scale anisotropic stress produced by matter. 

In order to evaluate the large-scale slip constraint in Equation (\ref{ls1}) it is useful to define an effective fluid, which we do by writing
\begin{equation} \label{eff}
G_{\mu \nu} = 8 \pi T^{\rm eff}_{\mu \nu} \, ,
\end{equation}
where $T^{\rm eff}_{\mu \nu}$ is the energy-momentum tensor of our `effective fluid'. Care should be taken to not interpret $T^{\rm eff}_{\mu \nu}$ as an actual fluid, though it can (and generically will) contain terms that correspond to real fluids in the space-time. 

We can compute the left-hand side of Equation (\ref{eff}) in perturbed Minkowski space, which, upon using the PPNC expressions for the gravitational potentials that appear, gives the following effective fluid components in expanding coordinates on small scales:
\begin{eqnarray} \label{rhoeff}
\rho^{\rm eff} &=& \gamma \, \rho - \frac{\gamma_c}{4 \pi} \\
p^{\rm eff} &=& \frac{(\alpha-\gamma)}{3} \rho - \frac{1}{12\pi} \left(\alpha_c-\gamma_c \right) \\
q^{\rm eff}_i &=& \gamma \rho v_i + \frac{1}{4 \pi} \left( \frac{\dot{\gamma}}{\gamma} + {H} \frac{(\alpha-\gamma)}{\gamma} \right) \, \hat{\Psi}_{,i} \, . \label{qeff}
\end{eqnarray}
It can be seen that the cosmological parameters $\{\alpha_c, \gamma_c\}$ contribute to $\rho^{\rm eff}$ and $p^{\rm eff}$, but not to $q^{\rm eff}_i$. This is as expected, as $\{\alpha_c, \gamma_c\}$ are dark energy parameters that affect the dynamics of the homogeneous background but not the perturbations {  (see e.g. \cite{ma})}.

One may note that the effective anisotropic stress, $\Pi^{\rm eff}_{ij}$, is missing from the list above. This is because it is derived from the slip equation, which has yet to be formulated. Using Equation (\ref{eff}), we expect this quantity to be given by
\begin{equation}
\Pi^{\rm eff}_{ij} = \frac{1}{8 \pi} \big( \hat{\Phi}-\hat{\Psi} \big)_{,\langle ij \rangle} \, ,
\end{equation}
where we have again used expanding coordinates. From Equation (\ref{ls1}) we can see that the large-scale slip will be governed by this small-scale effective anisotropic stress. From the discussion above, and in keeping with \hyperlink{assumption_1}{Assumption 1} from the Introduction, we assume that the cosmological parameters $\{\alpha_c, \gamma_c\}$ do not directly contribute to $\Pi^{\rm eff}_{ij}$. 

Returning to the effective energy density, we note that we are at liberty to split it into a part that is associated with matter, $\rho^{\rm eff,m}$, and a cosmological part that depends only on dark energy, $\rho^{\rm eff,d}$, such that
\begin{eqnarray} \label{rhosplit}
&&\rho^{\rm eff}=\rho^{\rm eff, m}+\rho^{\rm eff, d}
\end{eqnarray}
where
\begin{eqnarray}
&&\label{rhosplit2} 4\pi \, \rho^{\rm eff, m} = 4 \pi\, \gamma \, \rho - \frac{\partial \gamma_c}{\partial \rho} \, \rho 
\qquad {\rm and} \qquad 4\pi\, \rho^{\rm eff, d} = - {\gamma_c} + \frac{\partial \gamma_c}{\partial \rho} \, \rho \,.
\end{eqnarray}
Here, the terms involving a derivative with respect to $\rho$ account for any part of $\gamma_c$ that may come from the matter directly, as might happen, for example, in a simple scalar-tensor theory where $\varphi$ is sourced by the trace of the matter energy-momentum tensor. {  These definitions then result in $\rho^{\rm eff, m} \propto \rho$ and $\rho^{\rm eff, d}$  independent of $\rho$. We note that} the partial derivatives {  in these equations} should be taken after using the appropriate evolution equations to replace any quantities with time derivatives. Substituting Equation (\ref{rhosplit2}) into Equation (\ref{rhosplit}) then precisely recovers Equation (\ref{rhoeff}).

Similarly, the effective pressure can be split into matter-dependent and dark energy dependent parts as
\begin{equation}
p^{\rm eff}=p^{\rm eff, m}+p^{\rm eff, d} \, ,
\end{equation}
where
\begin{eqnarray}
&&4 \pi \, p^{\rm eff, m} = \frac{4\pi}{3}(\alpha-\gamma) \rho - \frac{\rho}{3} \frac{\partial}{\partial \rho}(\alpha_c-\gamma_c) \\
&&4\pi \, p^{\rm eff, d} = - \frac{1}{3} (\alpha_c-\gamma_c) +\frac{\rho}{3} \frac{\partial}{\partial \rho}(\alpha_c-\gamma_c)\, ,
\end{eqnarray}
and where we have again allowed for the possibility that the cosmological parameters, $\alpha_c$ and $\gamma_c$, can contain terms proportional to the matter energy density, $\rho$.

Unlike $\rho^{\rm eff}$ and $p^{\rm eff}$, the effective energy flux density $q^{\rm eff}_i$ in Equation (\ref{qeff}) contains only terms that are directly dependent on the matter; either directly through the energy density of matter, $\rho$, or through the gravitational potential $\Psi$, which is related to the matter density perturbation via $\nabla^2 \Psi = -4 \pi \, \gamma \, \delta \rho\, a^2$. We therefore write
\begin{equation}
q^{\rm eff}_i = q^{\rm eff,m}_i = \gamma \rho v_i + \frac{1}{4 \pi} \left( \frac{\dot{\gamma}}{\gamma} + {H} \frac{(\alpha-\gamma)}{\gamma} \right) \, \hat{\Psi}_{,i} \, .
\end{equation}
The reason for this different behaviour seems intuitively clear: while $\rho^{\rm eff}$ and $p^{\rm eff}$ have parts at homogeneous and isotropic background order that are sourced by both matter and dark energy, the first-order perturbations to these quantities are inhomogeneous and therefore can be sourced only by matter. Correspondingly, $q^{\rm eff}_i$ has no background component, and so is not sourced by dark energy at all.

We presume that the same logic should also be a feature of the effective anisotropic stress, $\Pi^{\rm eff}_{ij}$, which also has no background part. That is, we presume that an homogeneous and isotropic dark energy fluid cannot behave like an effective fluid with non-zero anisotropic stress \footnote{For an example of one of the small number of studies that assumes the opposite, see e.g. Ref. \cite{mota}.}. On the other hand, it seems entirely plausible that an effective fluid derived from the inhomogeneous and anisotropic distribution of real matter could do exactly that. This is in keeping with \hyperlink{assumption_1}{Assumption 1} from the Introduction, and is the philosophy that we will use to directly construct $\Pi^{\rm eff}_{ij}$.

Having divided our effective fluid into parts that are due to matter and dark energy, we can now write energy conservation equations for each of these:
\begin{eqnarray} 
&&\dot{\rho}^{\rm eff, m} + \Theta (\rho^{\rm eff, m}+p^{\rm eff, m}) + \sigma^{\mu \nu} \Pi^{\rm eff, m}_{\mu \nu} = -\Theta\, Q \\ \label{effc}
&&\dot{\rho}^{\rm eff, d} + \Theta (\rho^{\rm eff, d}+p^{\rm eff, d})  = \Theta \, Q \, ,
\end{eqnarray}
where $Q$ is the expansion-normalized energy flow between the two effective fluids. We can absorb $Q$ into the definitions of the effective isotropic pressures, $p^{\rm eff, m}$ and $p^{\rm eff, d}$, by treating it as a bulk viscosity term and defining the bulk-viscous pressures
\begin{equation}
\tilde{p}^{\rm eff, m} \equiv {p}^{\rm eff, m} + Q     \qquad {\rm and} \qquad \tilde{p}^{\rm eff, d} \equiv {p}^{\rm eff, d} - Q \, . 
\end{equation}
These definitions mean that the two effective fluids decouple from each other, which allows us to treat the potentially anisotropic effective fluid associated with matter separately from the isotropic effective fluid associated with dark energy. 

For the bulk-viscous effective fluid associated with the matter fields we have
\begin{equation}
\nabla^2 \phi^{\rm m} = - 4 \pi \, \left( \rho^{\rm eff, m} + 3 \tilde{p}^{\rm eff, m} \right) \qquad {\rm and} \qquad \nabla^2 \psi^{\rm m} = - 4 \pi \, \rho^{\rm eff, m} \, .
\end{equation}
By taking the energy exchange term $Q$ to {  be independent of spatial position (which must be the case, as it comes from the homogeneous $\rho^{\rm eff, d}$)}, we find the solutions
\begin{equation} \hspace{-1cm} \label{ppm}
\bar{\phi}^{\rm m} =  \left( \alpha -\frac{1}{4\pi} \frac{\partial \alpha_c}{\partial \bar{\rho}} + \frac{3 Q}{\bar{\rho}} \right) \, \bar{U} 
\qquad {\rm and} \qquad 
\delta \phi^{\rm m} = \alpha \, \delta U
\end{equation}
together with
\begin{equation} \label{ppm2}
\hspace{-1cm}
\bar{\psi}^{\rm m} = \left( \gamma -\frac{1}{4\pi} \frac{\partial \gamma_c}{\partial \bar{\rho}} \right) \, \bar{U}
\qquad \hspace{1.2cm} {\rm and} \qquad 
\delta \psi^{\rm m} = \gamma \, \delta U \, ,
\end{equation}
for the parts of $\phi^{\rm m}$ and $\psi^{\rm m}$ corresponding to the homogeneous background and the perturbations, respectively. Here $U$ is a solution of the Newton-Poisson equation $\hat{\nabla}^2 U = - 4 \pi \, \rho$, and we have defined $\hat{\nabla}^2 \bar{U} \equiv - 4 \pi \, \bar{\rho}$ and $\hat{\nabla}^2 \delta U \equiv - 4 \pi \, \delta \rho$.

We then get from Equations (\ref{ppm}) and (\ref{ppm2}) the following expression for the anisotropic stress associated with this effective fluid:
\begin{equation} \hspace{-1cm}
8 \pi \, \bar{\Pi}^{\rm eff, m}_{ij} = (\bar{\phi}^{\rm m} - \bar{\psi}^{\rm m})_{,\langle ij \rangle} =  \left( \alpha - \gamma -\frac{1}{4\pi} \frac{\partial \alpha_c}{\partial \bar{\rho}} +\frac{1}{4\pi} \frac{\partial \gamma_c}{\partial \bar{\rho}} + \frac{3 Q}{\bar{\rho}} \right) \bar{U}_{,\langle ij \rangle}
\end{equation}
and $8 \pi \, \delta {\Pi}^{\rm eff, m}_{ij} = (\alpha-\gamma) \delta U_{,\langle ij \rangle}$, where $U_{,\langle ij \rangle} = \bar{U}_{,\langle ij \rangle} + \delta U_{,\langle ij \rangle}$ is the Newtonian tidal force tensor split into background and first-order perturbations. Using Equations (\ref{ppm2}) and (\ref{effc}) to eliminate $\bar{U}_{,\langle ij \rangle}$ and $Q$ respectively, we have
\begin{equation} \hspace{-1cm} \label{sig1}
8 \pi \, \bar{\Pi}^{\rm eff, m}_{ij} = - \left[ \ln \left(\gamma - \frac{1}{4\pi} \frac{\partial \gamma_c}{\partial \bar{\rho}} \right)\right]^{\prime} \, \bar{\psi}^{\rm m}_{,\langle ij \rangle}  = 
\frac{d}{dt} \left[ \ln \left(\gamma - \frac{1}{4\pi} \frac{\partial \gamma_c}{\partial \bar{\rho}} \right)\right] \, \bar{\sigma}_{ij}
\, ,
\end{equation}
where we have made use of the integrability condition (\ref{intcon}), and in the last equality have imposed \hyperlink{assumption_1}{Assumption 1} to identify $\bar{\psi}_{,\langle ij \rangle} = -H\, \bar{\sigma}_{ij}$. 

Using Equation (\ref{ls1}) from Section \ref{sec:bianchi}, we can now see that the scalar part of the large-scale slip can be written as
\begin{equation} \hspace{-1cm} \label{larges}
\boxed{
\Phi - \Psi = - \frac{\hat{\gamma}'}{\hat{\gamma}} \, (\Psi - c_1) 
}\, ,
\end{equation}
where we have defined $4\pi \,\hat{\gamma} \equiv 4 \pi \, \gamma - {\partial \gamma_c }/{\partial \rho}$. {This relationship is valid for a homogeneous mode, as it comes from considering a homogeneous Bianchi cosmology as a perturbation of Friedmann. As such, it must correspond to the large-scale (super-horizon) limit of cosmological perturbations in our theory-independent approach. The result in Equation (\ref{larges}) can be seen to be significantly different to the small-scale (sub-horizon) limit given in Equation (\ref{ssslip}), which is that one that should be used to describe non-linear structures. In deriving this equation we have once again} taken the scalar part of the shear to be  $\bar{\sigma}_{ij} = \bar{v}_{,\langle ij \rangle}$ and used that $H v = c_1-\Psi$ on super-horizon scales. Here we have identified the scalar part of the large-scale velocity, $v$, with the homogeneous part of the small-scale velocity potential $\bar{v}$, and the homogeneous part of the scalar potential $\bar{\psi}$ with the perturbations describing Bianchi I cosmologies discussed in Section \ref{sec:bianchi}. This is our equation for the large-scale slip, with $c_1$ being the constant that was {discussed below Equation (\ref{c1eq})}. An alternative expression for this quantity, in terms of derivatives of PPNC parameters with respect to $H$, can be found in Appendix A.

\section{Example: Scalar-Tensor Theories of Gravity} \label{sec:st}

The discussion above has been in terms of a theory-independent parameterization of gravity. In order to examine the results that have been derived, we will now consider a specific class of theories of gravity; the canonical Bergmann-Wagoner \cite{bergmann, wagoner} scalar-tensor theories derived from the action
\begin{equation}
S= \frac{1}{16 \pi} \int d^4 x \sqrt{-g} \left[ \varphi R - \frac{\omega }{\varphi} \nabla^{\mu} \varphi \nabla_{\mu} \varphi - 2 \Lambda \right] \,,
\end{equation}
where $\varphi$ is a new fundamental scalar field, $\omega = \omega(\varphi)$ is a coupling function and $\Lambda=\Lambda(\varphi)$ is a cosmological term that can be thought of as the scalar field potential. By analysing the field equations derived from this action we can find the following expressions for the PPNC parameters:
\begin{eqnarray} \label{ag}
\hspace{-2cm} \alpha &=& \frac{2 (2+\omega)}{3+2 \omega} \frac{1}{\bar{\varphi}} \, , \hspace{1cm}
\gamma = \frac{2 (1+\omega)}{3+2 \omega} \frac{1}{\bar{\varphi}} \, ,\\
 \label{stac}
\hspace{-2cm} \alpha_c &=& - \omega \frac{\dot{\bar{\varphi}}^2}{\bar{\varphi}^2} + 3 H \, \frac{\dot{\bar{\varphi}}}{\bar{\varphi}} - \frac{8 \pi \, \bar{\rho}}{(3+ 2 \omega) \bar{\varphi}} 
+\frac{3 \dot{\bar{\varphi}}^2}{2 (3+2 \omega)} \frac{\mathrm{d} \omega}{\mathrm{d} \bar{\varphi}}
-\frac{3-2 \omega}{3+2 \omega} \frac{\Lambda}{\bar{\varphi}} + \frac{3}{3+2 \omega} \frac{\mathrm{d}\Lambda}{\mathrm{d} \bar{\varphi}} \, ,
\\ \label{stgc}
\hspace{-2cm}  \gamma_c &=& - \frac{\omega}{4} \frac{\dot{\bar{\varphi}}^2}{\bar{\varphi}^2} +  \frac{3H}{2} \frac{\dot{\bar{\varphi}}}{\bar{\varphi}} - \frac{4 \pi \, \bar{\rho}}{(3+2 \omega)\bar{\varphi}} 
-\frac{\Lambda}{2 \bar{\varphi}} \,,
\end{eqnarray}
{  where $\bar{\varphi}$ is the background (cosmological) value of the scalar field $\varphi$.} The reader may note that we have used the following evolution equation for the scalar field to remove highest-order derivatives of $\varphi$:
\begin{equation}
\ddot{\bar{\varphi}} + \Theta \, \dot{\bar{\varphi}} = \frac{8 \pi \, \bar{\rho}}{3+2 \omega}+ \frac{4 \Lambda}{3+2 \omega}  
-\frac{\dot{\bar{\varphi}}^2}{3+2 \omega} \frac{\mathrm{d} \omega}{\mathrm{d} \bar{\varphi}}
- \frac{2 \bar{\varphi}}{3+2 \omega} \frac{\mathrm{d}\Lambda}{\mathrm{d}\bar{\varphi}} \, .
\end{equation}
These expressions for the PPNC parameters have been checked to ensure compatibility with the equations of locally-rotationally-symmetric Bianchi I cosmologies in Brans-Dicke theory \cite{bi}, using the results from Section \ref{sec:bianchi}. 

\vspace{15pt}
\noindent
{\it Small-Scale Slip}
\vspace{5pt}

\noindent
After a post-Newtonian expansion, the field equations of these theories tell us that
\begin{equation}
\Phi - \Psi = \frac{\delta \varphi}{\bar{\varphi}} \, ,
\end{equation}
where $\varphi = \bar{\varphi} + \delta \varphi$ splits the scalar degree of freedom into a background part and a perturbation. At the same time, the leading-order part of the scalar field propagation equation on small scales is given by
\begin{equation}
\nabla^2 \delta \varphi = - \frac{8\pi}{(3+2 \omega)} \, \delta \rho\, a^2  = -4 \pi (\alpha-\gamma) \, \bar{\varphi} \, \delta \rho\,  a^2 \,.
\end{equation}
Using the small-scale Poisson equation for the spatial curvature perturbation,  $\nabla^2 \Phi = - 4\pi \gamma \, \delta \rho\,a^2$, then allows us to write  
\begin{equation}
\Phi - \Psi = \frac{\alpha- \gamma}{\gamma} \, \Psi \, ,
\end{equation}
which is precisely as expected from the theory-independent small-scale slip equation (\ref{ssslip}). The small-scale limit of our equations is therefore verified in this case.

\vspace{15pt}
\noindent
{\it Large-Scale Slip}
\vspace{5pt}

\noindent
For the large-scale limit of this class of theories, we can start by noting that the perturbed Friedmann equations of these theories give
\begin{equation} \label{stls1}
\left( \Phi - \Psi \right)_{,\langle ij \rangle} = - \frac{\dot{\bar{\varphi}}}{\bar{\varphi}} \, \sigma_{ij} \,,
\end{equation}
which has scalar part
\begin{equation} \label{stls2}
\Phi - \Psi =  - \frac{\dot{\bar{\varphi}}}{\bar{\varphi}} \, v    =\frac{\bar{\varphi}'}{\bar{\varphi}} \, (\Psi - c_1) \, ,
\end{equation}
where the prime again indicates a derivative with respect to the number of e-foldings. 

We can compare this to what we would find using the slip presented in Equation (\ref{larges}). In this case we need to calculate
\begin{equation}
\hat{\gamma} = \gamma - \frac{1}{4\pi} \frac{\partial \gamma_c}{\partial \rho} 
= \frac{2 (1+\omega)}{3+2 \omega} \frac{1}{\bar{\varphi}} - \frac{1}{4\pi} \left[ - \frac{4 \pi}{(3+2 \omega)\bar{\varphi}} \right] = \frac{1}{\bar{\varphi}} \, ,
\end{equation}
where after the second equality we have used the expression for $\gamma$ and $\gamma_c$ in Equations (\ref{ag}) and (\ref{stgc}). We can now see that Equation (\ref{larges}) does indeed recover the correct expression for the large-scale slip. We also note that the final equation in Appendix A also produces the correct result, after taking
\begin{equation}
\frac{\partial \gamma}{\partial H} =0 \qquad {\rm and} \qquad 
\frac{\partial \gamma_c}{\partial H} = \frac{3}{2} \frac{\dot{\bar{\varphi}}}{\bar{\varphi}} \, .
\end{equation}
We take this as verification of our approach. The transition between small and large scales is investigated in Appendix B.

\section{Numerical Results and Constraints} \label{sec:num}

In this section we will implement our new equation in a modified version of the Cosmic Linear Anisotropy Solving System (CLASS) \cite{class}, in order to demonstrate the validity of the approximations used in Reference \cite{ppnc5}. CLASS is a Boltzmann code that evolves linear perturbations around a Robertson-Walker geometry, for the purpose of calculating observables associated with large-scale structure and the cosmic microwave background (CMB). For deriving constraints we %follow Ref. \cite{ppnc5} and adopt a Markov Chain Monte Carlo (MCMC) technique, and use the MontePython code with a Metropolis Hastings algorithm and a jump factor of 2.1, together with 
use the Planck 2018 data %on TT, EE and TE correlations and lensing potentials 
\cite{planck1,planck2}. 

The free cosmological parameters in our analysis are the dimensionless cold dark matter and baryon densities $\omega_c$ and $\omega_b$, the Hubble rate $H_0$, the optical depth $\tau$, the primordial fraction of helium $Y_p$, and the amplitude and spectral index of scalar perturbations $A_s$ and $n_s$. %Nuisance parameters are treated using Gaussian priors \cite{planck1}, with flat priors on the cosmological parameters, and flat positive priors on the initial values of $\alpha$ and $\gamma$. 
In order to get constraints we postulate the following time-dependence for the gravitational parameters:
\begin{equation}
\alpha(a) = A \left(\frac{a_1}{a} \right)^n + B \qquad {\rm and} \qquad \gamma (a) = C \left( \frac{a_1}{a} \right)^n +D \, ,
\end{equation}
where the constants $A$, $B$, $C$ and $D$ are set by the values of $\alpha$ and $\gamma$ at reference times $a=a_1=10^{-10}$ (initially) and $a=a_0=1$ (today). The value of the power-law index $n$ can either be set to a test value or marginalized over, with tighter constraints being available with the former option \cite{ppnc5}. %These power-law forms are principally for simplicity, and that one may consider other functional forms or methods, as desired.

\begin{figure} 
\centering
\includegraphics[width=\textwidth]{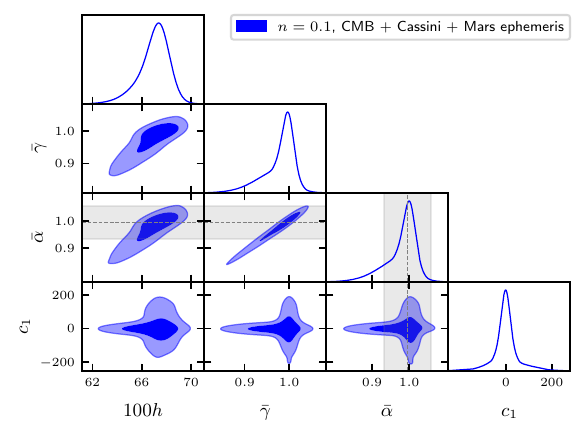}
\vspace{-1cm}
\caption{One and two-dimensional $0.3 \sigma$ smoothed posterior probabilities on the Hubble rate $100 \, h \equiv H_0/{\rm km}\hspace{0.1cm}  {\rm s}^{-1}  {\rm Mpc}^{-1}$, the parameters $\bar{\alpha}$ and $\bar{\gamma}$, and the constant $c_1$. The time-dependence of the gravitational parameters is a power-law with $n=0.1$, and the data used is from Planck \cite{planck1,planck2}, the Cassini probe \cite{cassini}, and the observations of the ephemeris of Mars \cite{mars}. The grey band shows the $1\sigma$ interval from the Mars ephemeris constraint on $\dot{\alpha}(a_0)$ converted into a Gaussian prior on $\bar{\alpha}$.
} \label{fig1a}
\end{figure}

In Figure \ref{fig1a} we present one and two-dimensional posteriors for the averages of the gravitational parameters,
\begin{equation}
\bar{\alpha} \equiv \frac{\int_{\ln a_1}^0 \alpha (a) \, {\rm d} \, \ln a}{\int_{\ln a_1}^0  {\rm d} \, \ln a}
\qquad {\rm and} \qquad
\bar{\gamma} \equiv \frac{\int_{\ln a_1}^0 \gamma (a) \, {\rm d} \, \ln a}{\int_{\ln a_1}^0  {\rm d} \, \ln a} \, ,
\end{equation}
together with those for $H_0$ and $c_1$. These constraints have been made using the value $\alpha(a_0)=1$, which is required by definition of Newton's constant, and where we have imposed the Gaussian priors $\dot{\alpha} (a_0) = 0.1 \pm 1.6 \times 10^{-13} {\rm yr}^{-1}$ and $\gamma (a_0) = 1 + (2.2 \pm 2.3) \times 10^{-5}$, coming respectively from observations of the ephemeris of Mars \cite{mars} and the Shapiro time delay measured by the Cassini space probe \cite{cassini}. We have also assumed that the radiation-dominated stage of the Universe's history proceeds unmodified, as it does in general relativity\footnote{{  Adding radiation to the PPNC approach is a work in progress \cite{ppncrad}, and is complicated by its absence as a matter source in most post-Newtonian treatments of astrophysical systems.}}, and that $\gamma_c$ is a constant and $\hat{\gamma} \simeq \gamma$. The evolution of $\alpha_c(a)$ is then determined by the integrability condition (\ref{intcon}). The PPNC coupling functions are smoothly interpolated in $k$ at each time interval, between their deep sub-horizon and super-horizon limits (for details see Ref. \cite{ppnc4}). We further assume vanishing spatial curvature, adiabatic perturbations, and two massless neutrinos and one with $m=0.06$ eV, with $N_{\rm eff} = 3.046$.

\begin{figure} 
\centering
\includegraphics[width=\textwidth]{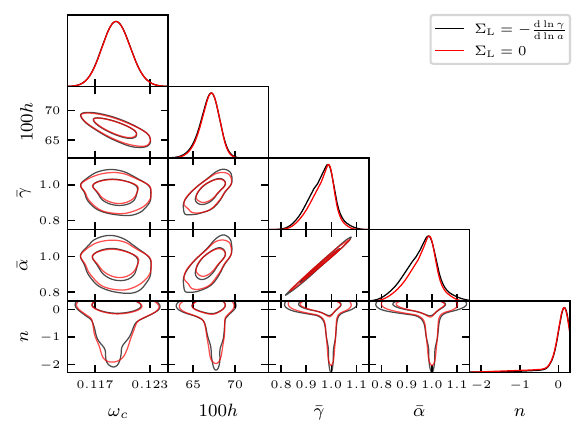}
\caption{A comparison of the one and two-dimensional constraints available using Eq. (\ref{larges}) for the large-scale slip, compared to assuming the general relativistic relation $\Phi_{\rm L} =\Psi_{\rm L}$, for the parameters $\omega_c$, $H_0$, $\bar{\alpha}$, $\bar{\gamma}$ and $n$.} \label{fig2}
\end{figure}

It can be seen from Figure \ref{fig1a} that the result $\alpha(a) \simeq \gamma(a)$, found in Reference \cite{ppnc5}, is also a result of the present analysis. We note a T-shaped degeneracy between $c_1$ and the other parameters being displayed. This happens as in the limit $\alpha=\gamma \rightarrow 1$ the gravitational theory reduces to general relativity. In this case the value of $c_1$ does not affect the physics, as expected from its multiplicative nature in Equation (\ref{larges}). The marginalized constraints on the values of $\bar{\alpha}$ and $\bar{\gamma}$ remain at the level of about 10\%, and are very much compatible with their general relativistic value of $\alpha=\gamma=1$. Finally, we compare our results with the slip from Equation (\ref{larges}) to the case where $\Phi=\Psi$ on large scales, as was assumed in Reference \cite{ppnc5}, and show the results in Figure \ref{fig2}. These results are for the special case $c_1=0$, which we again consider for simplicity. It can be seen that modifying the large-scale slip in a theoretically consistent manner has, in fact, very little effect on parameter constraints.

\section{Discussion} \label{sec:disc}

We have completed the PPNC formalism for dust and dark energy dominated cosmologies, by writing expressions for the slip equation on both large and small spatial scales. This was achieved on small scales by exploiting the direct isometry between perturbed Minkwoski and perturbed Robertson-Walker geometries on sub-horizon scales, and on large scales by considering homogeneous anisotropic cosmologies as perturbations of Friedmann cosmologies. This gave us Equation (\ref{ls1}), which links tidal stresses on large and small scales. We then obtained an expression for the anisotropic stress by treating the modified gravity phenomenology as being due to an effective fluid in section \ref{sec:direct}. Our results were derived under the assumption that only the gravitational field of matter contributes to the effective anisotropic stress.

Summarizing these results, we find that the slip equation in the PPNC formalism can be written as
\begin{equation} \label{result}
\boxed{\Phi - \Psi = \Sigma  \, \Psi + \mathcal{F}} \, ,
\end{equation}
where $\Sigma$ is defined such that
\begin{equation}
\lim_{k\rightarrow \infty} \Sigma =  \Sigma_{\rm S} \equiv \frac{\alpha-\gamma}{\gamma} \qquad {\rm and} \qquad \lim_{k\rightarrow 0}\Sigma = \Sigma_{\rm L} \equiv - \frac{\hat{\gamma}'}{\hat{\gamma}}  \, ,
\end{equation}
and where $\mathcal{F}$ is defined such that
\begin{equation}
\lim_{k\rightarrow \infty} \mathcal{F} =\mathcal{F}_{\rm S} \equiv 0 \hspace{25pt} \qquad {\rm and} \qquad \lim_{k\rightarrow 0} \mathcal{F} =\mathcal{F}_{\rm L} \equiv c_1 \, \frac{\hat{\gamma}'}{\hat{\gamma}}  \, ,
\end{equation}
{  where $k$ is the wavenumber of the perturbation and subscripts ${\rm S}$ and ${\rm L}$ denote quantities defined in the limits of small and large scales, respectively.} The function $\Sigma$ is reminiscent of the $\mathcal{G}$ that appears in the large-scale momentum constraint equation (\ref{q}), except here it has some dependence on the part of $\gamma_c$ that is a function of $\rho$ (in the form of the second term in $\hat{\gamma}$). The interpolation between the short and long wavelength regimes in this equation is investigated in Appendix A.

We also investigated the applicability of the equations presented above in the case of canonical scalar-tensor theories of gravity in Section \ref{sec:st}. We found that both Equations (\ref{effcon}) and (\ref{larges}) are applicable in this case, and that they reproduce precisely the expected results for this class of theories. This provides justification for the validity of our theory-independent approach. While the slip equations derived here are less straightforward than the rest of the PPNC set, they provide a crucial missing piece of this theory-independent framework, which would otherwise require extra phenomenological parameters that would not be linked to those that are familiar from the parameterized post-Newtonian approach to weak-field gravitational physics.

In the final part of this presentation, in Section \ref{sec:num}, we show the result of including our new equations into the CLASS code, and applying them to CMB observables. We find results that are both qualitatively and quantitatively close to those that were obtained by assuming that we could simply set $\Phi=\Psi$ on large spatial scales. This is a reassuring result, which indicates that the detailed analysis performed in Reference \cite{ppnc5} is not substantially altered by a more careful parameterization of the slip equation in the cosmologies of alternative theories of gravity.

The gravitational slip expression that we have derived here will make it possible to convert constraints on $\Sigma$ from cosmological large-scale structure data (such as weak lensing measurements from Euclid \cite{euclid}) into direct constraints on the PPN parameters themselves. As this result completes the full set of PPNC scalar equations, the framework can be applied directly to those datasets, thereby allowing tests of General Relativity without reference to any specific class of alternative theoretical models. Moreover, these cosmological tests can be combined directly with astrophysical measurements, allowing a single holistic test of gravity to be carried out across all those scales. In an upcoming work we will do precisely this, by combining Solar System tests with observations of both the CMB anisotropies and the baryon acoustic oscillations, in order to provide the most wide-ranging theory-agnostic tests of gravity to date \cite{ppnc_bao}. This will be followed in the future by further work interpreting large-scale structure data in the light of the PPNC framework.

\vspace{0.75cm}
\noindent {\bf Acknowledgements:} TA, TC and DBT acknowledge support from the Science and Technology Facilities Council (STFC, grant numbers ST/P000592/1 and ST/X006344/1), and well as assistance from the QMUL ITS Research team, and helpful conversations with Phil Bull. This research utilised Queen Mary's \href{http://doi.org/10.5281/zenodo.438045}{Apocrita HPC facility}, supported by QMUL Research-IT.

\section*{Appendix A: \hspace{0.25cm} An Alternate Expression for the Large-Scale Slip}\label{sec:appendix_a}

Some further information about the form of $\Pi^{\rm eff}_{ij}$ can be obtained directly from the other effective fluid quantities, by considering the Bianchi I cosmologies constructed in Section \ref{sec:bianchi}. Choosing a set of frames that are comoving with the dust in these space-times gives
\begin{equation}
\dot{\rho}^{\rm eff} + \Theta  \left( \rho^{\rm eff} + p^{\rm eff} \right) = - \sigma^{\mu \nu} \Pi^{\rm eff}_{\mu \nu} \, ,
\end{equation}
where $\Theta$ is the expansion scalar and $\sigma^{ij}$ is the shear tensor associated with the dust. Evaluating the left-hand side of this equation using equations (\ref{rhoeff})-(\ref{qeff}), and using the energy conservation equation for dust $\dot{\rho}+ \Theta \, \rho =0$, gives
\begin{equation}
- \sigma^{\mu \nu} \Pi^{\rm eff}_{\mu \nu}=  \dot{\gamma} \, \rho + \frac{1}{3} (\alpha-\gamma) \, \rho \, \Theta - \frac{\dot{\gamma}_c}{4 \pi} - \frac{1}{12 \pi} \left(\alpha_c+2 \gamma_c \right) \Theta  \equiv I\, .
\end{equation}
In the Friedmann limit, $\sigma \rightarrow 0$, this can be seen to give $I=0$, which is equivalent to the integrability condition (\ref{intcon}). 

If we now take $\Pi^{\rm eff}_{\mu \nu} = - \lambda \, \sigma_{\mu \nu}$, then we find $\lambda = \frac{1}{2} I \sigma^{-2}$, where $\sigma^2 \equiv \frac{1}{2} \sigma_{\mu \nu} \sigma^{\mu \nu}$. The quantity $\lambda$ in such an expression can be interpreted as the shear viscosity of the effective fluid. Taking the Friedmann limit in this case now gives
\begin{equation} \label{pieff}
\Pi^{\rm eff}_{\mu \nu} = - \frac{1}{2} \sigma_{\mu \nu} \, \lim_{\sigma^2 \rightarrow 0}    \frac{I}{\sigma^2} \,.
\end{equation}
Suppose we now introduce \hyperlink{assumption_2}{Assumption 2} from the Introduction, so that the PPNC parameters $\{\alpha,\gamma,\alpha_c,\gamma_c\}$ can be functions of the expansion scalar $\Theta$, but are not directly dependent on the shear $\sigma_{\mu \nu}$. Then, we can write Equation (\ref{pieff}) as
\begin{equation} \label{pieff2}
\Pi^{\rm eff}_{\mu \nu} = \left( \rho \frac{\partial \gamma}{\partial \Theta} - \frac{1}{4\pi} \frac{\partial \gamma_c}{\partial \Theta} \right) \sigma_{\mu \nu} \, ,
\end{equation}
which follows from the Raychaudhuri equation $\dot{\Theta} = - 2 \sigma^2 + \dots$, where dots denote additional terms that are not important here. The partial derivatives here indicate derivatives with respect to $H$, after evolution equations have been used to remove terms with time derivatives in expressions for $\gamma$ and $\gamma_c$. We anticipate Equation (\ref{pieff2}) to be more useful than (\ref{pieff}), as it requires expressions for $\gamma$ and $\gamma_c$ in the background Friedmann geometry only, and does not require any consideration of Bianchi I cosmologies\footnote{This method of determining $\Pi^{\rm eff}_{\mu \nu}$ does, however, rely on knowing the functional dependence of $\gamma$ and $\gamma_c$ on the expansion scalar $\Theta$, which will require knowledge of the underlying theory being tested.}.

Using the expressions from Section \ref{sec:bianchi}, we can now write the scalar part of the large-scale anisotropic stress as
\begin{equation} \label{effcon}
\Phi-\Psi = - \frac{1}{H} \left( \frac{8\pi}{3} \, \rho \, \frac{\partial \gamma}{\partial H} - \frac{2}{3} \frac{\partial \gamma_c}{\partial H} \right) \, (\Psi - c_1) \, ,
\end{equation}
where we have taken $\Theta = 3 H$, and where the scalar part of the shear is given by the scalar velocity potential as $\sigma_{ij} = v_{,\langle i j \rangle}$, such that $H v=c_1-\Psi$ (with $c_1$ here being the constant discussed at the end of Section \ref{sec:bianchi}) \cite{ppnc5}. The result in Equation (\ref{effcon}) is valid for any model of modified gravity and/or dark energy in which the dark energy field does not directly produce shear. For more general theories one would require the use of Equation (\ref{pieff}).

{We note that the constant $c_1$ appears to play different roles in the different expansions that we have used to construct our theory-independent version of cosmological perturbation theory. That is, in the comparison between different homogeneous cosmologies, which we used to obtain the large-scale of cosmological perturbations, it comes in as a freedom to rescale the spatial coordinates of homogeneous cosmologies (as discussed below Equation (\ref{c1eq}). In the resultant cosmological perturbation theory equations, however, it is best thought of as an initial condition that sets the amplitude of perturbations (see, e.g., Ref. \cite{ppnc5}). These are actually two different aspects of the same physics, as super-horizon perturbations with different amplitudes can be understood as homogeneous cosmologies with re-scaled spatial sections.}

\section*{Appendix B: \hspace{0.25cm} Scale dependence}\label{sec:appendix_b}

\begin{figure}[ht]
\centering
\includegraphics[width=0.7\textwidth]{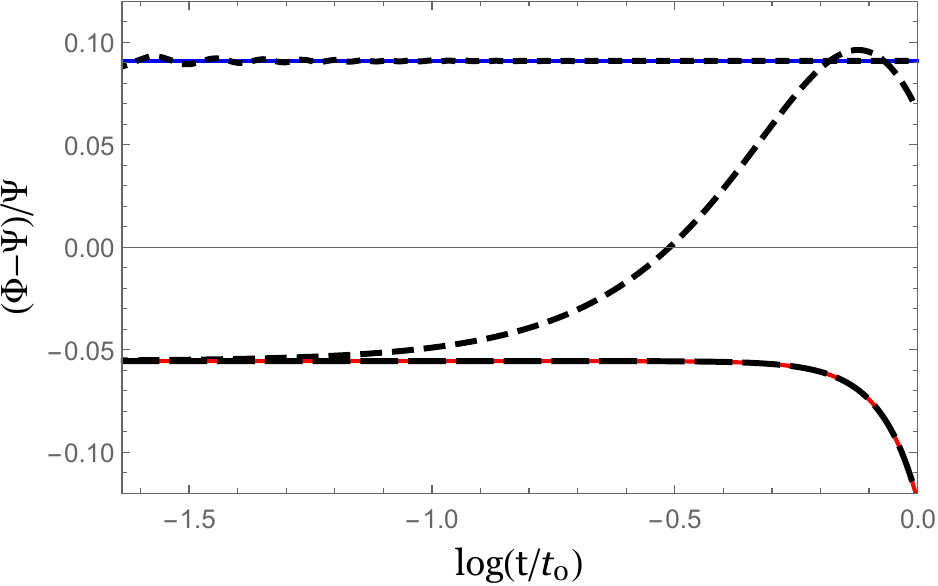}
\caption{Evolution of the gravitational slip on small ($k=6 \pi \, H_0 \times 10^2$), medium ($k=6 \pi \, H_0$) and large ($k=6 \pi \, H_0 \times 10^{-2}$) spatial scales, displayed as dotted, dashed and long-dashed lines respectively, for a scalar-tensor theory with $\omega=10$ and $\Omega_{\Lambda}=0.7$. The blue and red lines show the small and large-scale limits prescribed by Equation (\ref{result}) for the cases $k=6 \pi \, H_0 \times 10^2$ and $k=6 \pi \, H_0 \times 10^{-2}$.}
\label{figA}
\end{figure}

The results summarized in Section \ref{sec:disc} are shown graphically in Figure \ref{figA}, for the case of a scalar-tensor theory with constant $\omega=10$ and $\Omega_{\Lambda}=0.7$ (chosen so that it demonstrates a significant departure from general relativity in the slip parameter, rather than for its realism or observational viability). The blue and red lines in the plot in this figure are the predictions for the small and large-scale gravitational slip given by Equation (\ref{result}). The dotted lines show the result of directly integrating the perturbation equations for small-scale perturbations in these theories, and matches very closely with the blue line (the small wiggles result from the scalar field settling down to its small-scale limit). The long-dashed line shows the result of integrating the perturbation equations on large scales, and clearly reproduces the results of the red line well. In between these two results, we have a mode that crosses the horizon scale during the plotted range of $t$, and which we have shown as a short-dashed line. This curve starts at the large-scale limit (i.e. the red line) at early times, and ends up oscillating around the small scale-limit (i.e. the blue line) after it crosses the horizon (with further damped oscillations occurring around the small-scale limit if this line is continued into the future). This plot demonstrates our proposed expressions for the gravitational slip work well in the appropriate limits, agreeing with the numerical solutions as required, and permitting a sensible interpolation between the small and large-scale limits \cite{ppnc4, trans}.

\hspace{1cm}

\newpage

\phantom{a}

\vspace{-1cm}
\section*{References}

\bibliographystyle{unsrt}

\end{document}